# Feature Articles
## Articles de Fond

## The Plaskett Lecture: Star Formation in the Perseus Molecular Cloud

*Helen Kirk*
*NSERC PDF, Harvard Smithsonian Center for Astrophysics*
*hkirk@cfa.harvard.edu*

## Foreword

This article is based on my Plaskett Medal Lecture given at the annual CASCA meeting in Halifax, in May 2010. The Plaskett Medal is awarded to "the Ph.D. graduate from a Canadian university judged to have submitted the most outstanding doctoral thesis in astronomy or astrophysics in the preceding two calendar years." The lecture summarized my research during my thesis, as does this article. Contributions from Canadian facilities are highlighted where appropriate.

## Abstract

Large-scale surveys of the Perseus molecular cloud have provided many clues as to the processes occurring during star formation. Here, analysis of both column density maps and kinematic data (maps and pointed data) are discussed and compared with predictions from simulations. Results include a column density threshold for the formation of dense star-forming cores and that the dense cores are quiescent within their local environment, while the molecular cloud as a whole has turbulent motions that are dominated by large-scale modes. Some of these results have already been used to constrain models of star formation, and the others can be included as future tests of the models. The next few years of star formation research promises to provide exciting advances to the field, particularly with the Gould Belt Legacy Surveys in progress at several facilities, including the James Clerk Maxwell Telescope (JCMT).

## 1. Introduction

Stars form within large complexes of gas and dust (at a ratio of ~100:1) known as molecular clouds. Molecular clouds were originally discovered as dark patches on the sky where no stars were visible. These "dark clouds" were first catalogued by Herschel (1785) and Barnard (1913); at the time, it was not known whether the dark patches represented an absence of stars or the obscuration of starlight by an unknown source. It took the advent of the radio telescope to solve the question – observations of the dark cloud in Orion revealed emission from the gas-phase molecule CO (carbon monoxide) (Wilson, Jefferts, & Penzias 1970), confirming the presence of material within them. CO is actually the second most abundant gas-phase species in a typical cloud. $H_2$ (molecular hydrogen) is the dominant species by a factor of at least $10^5$. $H_2$ is not, however, observable under usual molecular-cloud conditions, so CO is often used as a proxy. Also found with $H_2$, CO, and other gas-phase molecules are dust grains, roughly micron-sized solid particles, which are responsible for the stellar obscuration.

We now know that stars typically form in these dark clouds, giant complexes of molecular gas and dust that span tens of parsecs (~30 light-years) and contain ~10,000 times the mass of material in our own Sun. Molecular clouds are hierarchically structured, and most of the mass (and area or volume) of the cloud is at low (column) densities, with only a small fraction in more condensed structures. The most compact structures are dense cores, the birthplace of stars. The large-scale clouds have densities around 100 particles per $cm^3$ and column densities of ~$10^{21}$ particles per $cm^2$, while the associated dust obscures the background starlight by an $A_V$ (visual extinction) of a few magnitudes. The cores have densities in excess of $10^5$ $cm^{-3}$ and column densities corresponding to tens or hundreds of $A_V$. To put this in perspective, even these dense cores are substantially more rarefied than the Earth's atmosphere!

The field of star-formation study is still young, with a mere 40 years of radio observations. There is still much to learn, despite the tremendous progress already made. This article focusses on observations connecting the dense cores to their larger environment. These new results provide a valuable way to learn about the processes that shape the formation and evolution of both the molecular cloud as a whole and the dense cores and stars forming within them.

A first look at a molecular cloud is puzzling. At the typical density and temperature for the cloud as a whole, thermal pressure can only prevent gravitational collapse for at most a few hundred solar masses worth of material. Molecular clouds contain hundreds of times more material than this, so how could they have formed that way without immediate collapse? There are two possible solutions. Gravitational collapse could be inhibited by additional physical processes. Alternatively, each cloud may not be a single well-defined entity – some parts may be collapsing while other parts are beginning to form, so that the "cloud" we observe stays visible for a much longer time period than any individual piece within it. Likely both of these are true to some extent. In this context, a variety of physical processes have been proposed for star formation. A few of the most popular ones include magnetic fields (*e.g.* Shu, Adams, & Lizano 1987), supersonic turbulent motions (*e.g.* MacLow & Klessen 2004), "global gravitation" (*e.g.* Burkert & Hartmann 2004), and "triggering" (*e.g.* Elmegreen 1998), all of which are briefly outlined below.

Magnetic fields primarily influence the behaviour of ionized atoms or molecules within the cloud, inhibiting their motion perpendicular to the field lines. If the field is strong enough, then even neutral species have their motions reduced perpendicular to the field lines, slowed by frequent collisions with the motion-restricted ions. In this case, mass can only be concentrated into denser structures slowly, by the neutral species "drifting" past the ions in a process known as ambipolar diffusion. With a strong enough magnetic field, the timescale for gravitational collapse can be increased by a factor of ten or more.

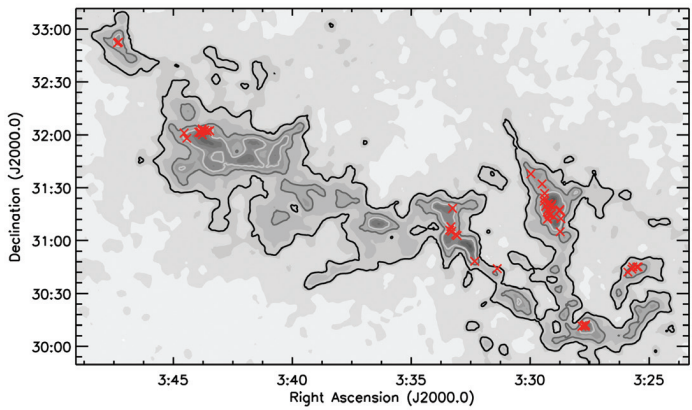

*Figure 1 — An extinction map of the Perseus molecular cloud. Contours are shown at $A_v$ of 3, 5, and 7 (dark to light lines). The locations of the dense cores identified are overlaid as red crosses. The SCUBA map in which the dense cores were identified spans a contiguous area covering beyond the black contour. Clearly, the dense cores tend to be found in regions of high extinction, despite the substantial coverage outside of these regions. Adapted from Kirk, Johnstone, & Di Francesco (2006).*

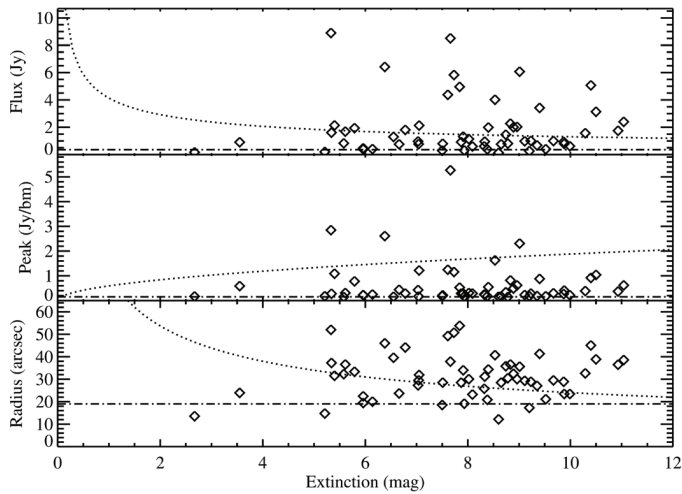

*Figure 2 — Dense core properties as a function of the local extinction (large-scale column density) environment. From top to bottom: the total flux, peak flux, and radius of the dense cores. The diamonds show the observations, while the dotted line shows the behaviour of a Bonnor-Ebert sphere in varying pressured environments (low pressure corresponds to low column density or $A_v$). The dash-dotted line shows the approximate observational limits. Dense cores similar to those in regions of high extinction would be easily detectable at lower extinctions. Adapted from Kirk, Johnstone, & Di Francesco (2006).*

Turbulent models come in a variety of forms, but in the most general picture, there are large-scale supersonic flows of material within molecular clouds. On the cloud-wide scale, these motions act to increase the effective temperature of the gas, allowing a larger amount of material to be stable to immediate gravitational collapse. Where the flows collide, the density is temporarily enhanced and can lead to the formation (and subsequent gravitational collapse) of dense cores. There are currently two main contending processes for how dense cores accrete their mass. The first is competitive accretion, in which a clustered distribution of protostellar "seeds" compete with each other for material within the larger cluster's gravitational potential. "Seeds" near the centre tend to gain more mass than the ones at the outskirts; interactions between the *seeds* play a major role in their evolution (*e.g.* Bonnell *et al.* 2001). The second scenario, monolithic collapse, posits that each dense core provides a self-sufficient reservoir for one or a small group of forming stars, so the mass that each star (or group) draws from is distinct; interactions between protostars are much less important (*e.g.* Tan, Krumholz, & McKee 2006).

The global gravitation model is related to the turbulent scenario above. The main distinction is that the turbulence is explicitly generated by gravity. Here, gravity acts on the large-scale structure of the cloud during its formation, inducing turbulent motions that are continuously driven by the large-scale evolution of the cloud. Material near the cloud edges where the curvature is higher tends to collapse first (in 2-D, finger-like regions coalesce before pancake-like regions), and the non-uniform motions induced by the evolution of the uneven "edges" of the molecular cloud percolate inwards, driving non-thermal motions in the cloud interior.

Triggering is a general term that encompasses a number of scenarios. Some triggering events start with a pre-existing stable structure, such as a dense core. The structure is then "triggered" to undergo gravitational collapse through an increase of pressure in its environment, typically caused by a nearby young massive (O or B) star "turning on" and increasing the radiation pressure in the local environment. Other triggering events can both create the structure and induce the subsequent evolution. An example of this is a supernova explosion – the shock wave from the explosion radiates outwards, sweeping up material in its path and collecting it into a ring or shell. Eventually, enough mass is collected that the ring/shell becomes gravitationally unstable and then fragments into a series of dense cores that could subsequently form stars.

These proposed processes have a major (and differing) impact on the large-scale evolution of molecular clouds, so determining their importance requires cloud-wide observations. Earlier observations focussed on regions where star formation was known to occur, but over the last several years, instrumentation has improved to the point where large-scale surveys are possible at a variety of wavelengths. Different wavelengths and different tracers are sensitive to complementary features of the star formation process; coordinated campaigns are vital for obtaining the full picture. With unbiased surveys, statistical measures of the properties of the clouds and the dense cores within them have the potential to rule out or place limits on the importance of the processes discussed above.

The Perseus molecular cloud is one of the best-studied molecular clouds to date, and I will highlight results from a coordinated survey of Perseus in which I was involved. Many of the observations were obtained under the auspices of the COM-PLETE (COordinated Molecular Probe Line Extinction Thermal Emission) survey. COMPLETE was one of the earliest large multi-telescope, multi-wavelength star formation endeavours, focussing primarily on two nearby star-forming regions, Perseus and Ophiuchus (Goodman 2004, Ridge *et al.* 2006). Large-scale surveys are now underway at several telescopes (including the JCMT, *Spitzer*, and *Herschel*) covering many nearby molecular clouds, so the results and analyses discussed here offer a preview into the insights these new surveys will provide. Where possible, comparisons with predictions from the models described above are discussed. The majority of simulations to date focus on turbulence; few include magnetic fields, and only recently have predictions been made in the observational domain. Global gravitational simulations do not yet have the resolution for comparison with dense-core observations (see *e.g.* Heitsch *et al.* 2008), and so will not be discussed further here.

## 2. Observations

The Perseus molecular cloud is a relatively nearby (250 pc/800 ly) star-forming region, and has properties mid-way between more extreme examples such as Taurus and Orion. Perseus shows more active and clustered star formation containing a mix of low and intermediate-mass stars, while Taurus is a very quiescent cloud that appears to be only forming a dispersed population of low-mass stars. Yet Perseus appears almost Taurus-like in comparison to the Orion molecular cloud, where a large population of high- and low-mass stars is forming dynamically in several large clusters. Across the galaxy, most stars appear to form in clusters (Lada & Lada 2003), and those in Perseus are more amenable to study than in Orion, as they are in smaller, sparser clusters with less source confusion. The entire Perseus cloud spans several square degrees on the sky, and has been mapped in a variety of tracers, some of which are described below.

### 2.1 Column Density Structure

The column density within molecular clouds is best measured using two complementary techniques. The large-scale column density structure is inferred using deep stellar catalogues. The dusty component of molecular clouds blocks (or extincts) the light from background stars, preferentially at shorter (bluer) wavelengths; areas with fewer and redder stars thus have a larger column density of material in front of them. One technique that has been applied to Perseus is NICER (Near Infrared Colour Excess - Revisited; Ridge *et al.* 2006). NICER (Lombardi & Alves 2001) uses observations at three near-infrared wavelengths to measure the stellar reddening and number density, taking advantage of the fact that most stars have similar intrinsic colours in the infrared and the extinction is less than at optical wavelengths. The derived dust column density from NICER or similar techniques is dependent only on the simple physics of dust grains scattering light – the temperature of the dust, for example, has no effect on the measure. The small-scale structure, however, cannot be probed with this method – the statistical measure of stellar reddening requires averaging over multiple stars; without very deep, targeted observations, a resolution of a few arcminutes is usually the best obtainable. At the distance of Perseus, dense cores are much smaller than this resolution, and cannot be probed with this method (for one region where "extinction mapping" has been successfully applied to the scale of cores, see Lombardi *et al.* 2006).

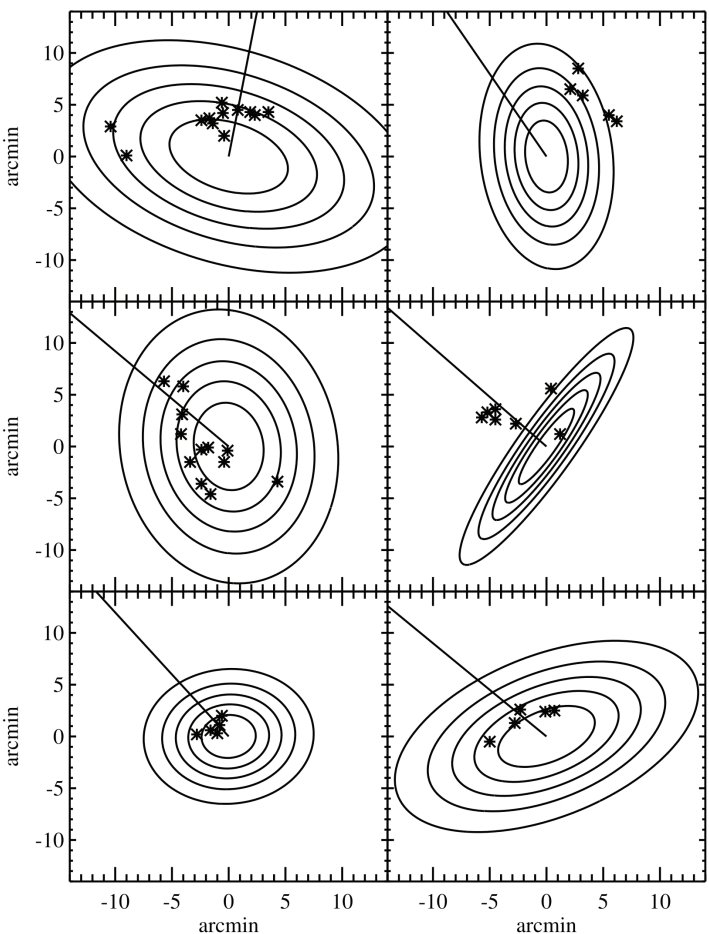

*Figure 3 — A snapshot of the extinction peaks that harbour the most dense cores, showing the tendency of the dense cores to be offset from the peak of the extinction. The contours show a Gaussian model of the extinction, while the asterisks show the locations of the dense cores. The solid line points in the direction of 40 Per, which other work has suggested is triggering star formation in Perseus. There is reasonable qualitative agreement between the offset of the dense cores and the direction of 40 Per. Adapted from Kirk, Johnstone, & Di Francesco (2006).*

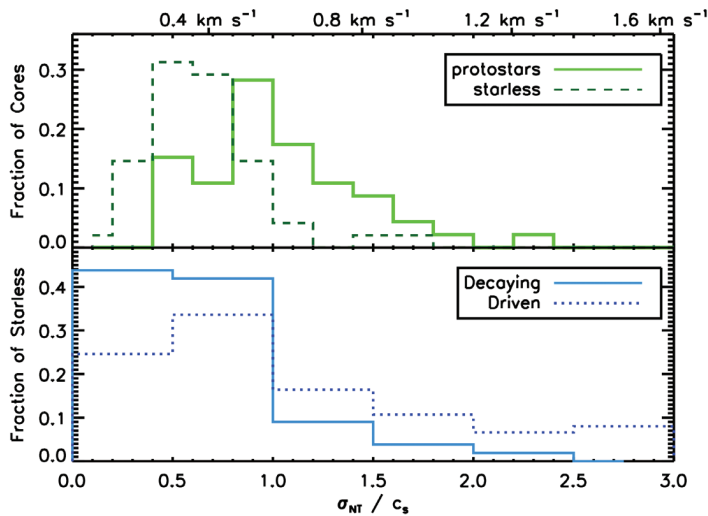

*Figure 4 — The non-thermal motions within dense cores. The top panel shows the observed ratio of the non-thermal linewidth to the sound speed, split into starless and protostellar cores (dashed and solid lines). The bottom panel shows predictions for starless cores from two recent simulations by Offner et al. (2008), where the turbulence was allowed to decay (solid line) or was continuously driven (dotted line). The final bin for the driven turbulence simulation encompasses all data up to a ratio of 6. Clearly, the driven turbulence simulation is highly inconsistent with our observations of starless cores. Adapted from Kirk, Johnstone, & Tafalla (2007).*

The small-scale column density distribution is measured by observing the thermal emission from the dust, typically using a submillimetre telescope, such as the James Clerk Maxwell Telescope (JCMT) in Hawaii, in which Canada is a major partner. The JCMT offers a resolution of 10 to 20 arcseconds, enough to (barely) resolve the dense cores. The thermal emission of dust can be used to estimate the column density, although care must be taken to consider the effect of temperature and varying dust-grain properties. The large-scale dust column density distribution cannot be measured by ground-based submillimetre facilities – the telescopes essentially make difference measures across the sky (necessitated by observing conditions), and so are insensitive to large-scale structure. In Perseus, the small-scale column density structure was mapped at the JCMT using SCUBA (Submillimetre Common-User Bolometer Array) primarily by Hatchell *et al.* (2005), with additional observations in Kirk, Johnstone, & Di Francesco (2006, hereafter Paper I). A slightly lower-resolution map was later made at the nearby Caltech Submillimeter Observatory (CSO) by Enoch *et al.* (2006).

### 2.1.1 Dense Core Extinction Threshold

In Paper I, we identified and analyzed the dense cores found in the JCMT SCUBA map, and then compared the locations of the dense cores within the large-scale cloud structure seen in the NICER extinction map. Figure 1 shows the extinction map, with contours at 3, 5, and 7 magnitudes of extinction and the locations of the dense cores overlaid as red crosses. Even at a quick glance, it is easy to see that the dense cores are not randomly distributed throughout the cloud – they lie only in regions of high extinction, despite the fact that the survey included a large area of the cloud at lower extinctions.

Modelling was required to test whether the lack of detection of cores at low extinction was due to observational sensitivity or whether it represented a true dearth of cores. Previous work (*e.g.* Johnstone *et al.* 2000) showed that dense cores can be roughly represented as Bonnor-Ebert (BE) spheres: isothermal, pressure-bounded spheres where self-gravity is balanced by thermal pressure. Using the BE sphere model, the observable properties of a dense core were predicted at different column densities (or extinctions), assuming that at lower column densities, the bounding pressure on the cores is smaller, as there is less overlying cloud material weighing on them. At lower column densities, therefore, the same core would be larger, less peaked, and able to support a larger amount of material against gravitational collapse. Figure 2 shows the predicted behaviour of core properties as a function of cloud extinction (dotted lines). The observed core properties are indicated by the diamonds, while the observational sensitivity is shown by the dash-dotted line. Clearly, dense cores similar to those found in the highest extinction parts of the cloud were detectable in the lower extinction parts; their absence indicates they were not able to form there.

A similar threshold for dense cores has also been found in several other star-forming regions, including Taurus (Onishi *et al.* 1998), Ophiuchus (Johnstone, Di Francesco, & Kirk 2004), and later Perseus through independent data and analysis (Enoch *et al.* 2006). A column density threshold for dense core formation was actually predicted by McKee (1989) for a magnetically dominated star-formation scenario. The basic premise of the paper was that dense cores can only coalesce in a reasonable amount of time in the very interior of molecular clouds. There, fewer ions exist to impede the formation of dense cores, as magnetic field lines inhibit the motion of ions across them. In the cloud exterior, the impinging UV radiation field ionizes too large a fraction of the cloud material, effectively hindering the motion of both ionized and non-ionized material. The extinction threshold predicted by McKee (1989), an $A_V$ of 4 - 8 magnitudes, agrees with the observations. Simulations of turbulence-dominated or global gravity-dominated formation have not yet had sufficient dynamic range to test for this property.

### 2.1.2 Offset Core Locations

Upon a more thorough examination of Figure 1, a second property becomes apparent – while the dense cores were found in regions of higher column density, they appeared to be offset from the peaks of the extinction, in a preferential direction. Previous work by Walawender *et al.* (2004) showed that 40 Per, a nearby young B star, appeared to be shaping the cometary L1451 cloud (which lies below the bottom right-hand corner of Figure 1). The core offsets from the extinction peaks we found in Paper I were consistent with this triggering scenario. Figure 3 shows the six extinction peaks that harbour the densest cores. Contours indicate a Gaussian fit to the extinction, while the dense cores are shown as asterisks. The solid line points in the direction of 40 Per, showing that most of the offsets are in qualitative agreement with a triggered formation scenario.

### 2.2 Kinematics

Emission from gas-phase species within molecular clouds provides complementary information about the star-forming environment. Since clouds and cores are cold, often at temperatures around 15 K (-258 °C), the lowest few rotational energy level transitions in molecules (typically at radio or millimetre / submillimetre wavelengths) are best suited for detections. Since $H_2$ is often unobservable in this regime, trace species, such as CO, are commonly employed to study the cloud. Unlike the dust-emission observations, spectral observations provide velocity information allowing the measurement of both the bulk motion of the gas (through the wavelength of the centre of the emission line) and the level of motion between particles within the gas (through the width of the line). Chemistry and physical conditions affect which molecular transitions are observable, and are necessary to consider when interpreting observations. For example, CO and other carbon-bearing molecules tend to freeze out of the gas phase and onto dust grains at low temperatures and high densities, *i.e.* within dense cores (Tafalla *et al.* 2002). CO therefore provides an excellent tracer of the bulk cloud gas, but cannot be used to measure cores. Nitrogen-bearing molecules, such as $NH_3$ (ammonia) or $N_2H^+$ (diazenylium), on the other hand, trace the dense cores well, but not the bulk cloud gas, as their formation is inhibited by the presence of gas-phase CO. Different molecules therefore must be observed, depending on what is being studied.

In Kirk *et al.* (2007, hereafter Paper II), we surveyed the kinematics of the dense cores in Perseus with a pointed spectral survey of all of the dense cores identified in Paper I, as well as additional candidate dense cores, for a total of 157 targets. The survey included $N_2H^+$ (1-0) and $C^{18}O$ (2-1), which can be observed simultaneously at the IRAM 30-m telescope in Granada, Spain. $N_2H^+$ was used to probe the dense core, while $C^{18}O$ was used to probe the lower-density envelope surrounding the core; $C^{18}O$ is sensitive to a somewhat higher-density regime than its isotopologues $^{12}CO$ and $^{13}CO$. Cores were classified as those that had already formed a young central source (protostars), and starless cores, based on a prior analysis that combined the SCUBA data with near- to mid-infrared observations from *Spitzer* (Jorgensen *et al.* 2007).

### 2.2.1 Thermally Dominated Cores

Consistent with previous work (*e.g.* Jijina, Myers, & Adams 1999), we found (Paper II) that the dense cores tended to have narrow linewidths, showing that the cores were thermally dominated. Assuming a temperature of 15 K (similar to that later measured in the Perseus cores in $NH_3$ by Rosolowsky *et al.* (2008), the non-thermal portion of the linewidth, $\sigma_{nt}$, was roughly half as large as the local sound speed, $c_S$, for the starless cores, and roughly equal for the cores that have already formed a protostar. Our survey was the first to make such a characterization for a homogenous population of dense cores in a single cloud; previous surveys typically used only a handful of cores from any given cloud. With these data, statistically significant comparisons with simulations and models could be made. Figure 4 shows a comparison of the ratio of $\sigma_{nt}$ to $c_S$ for the

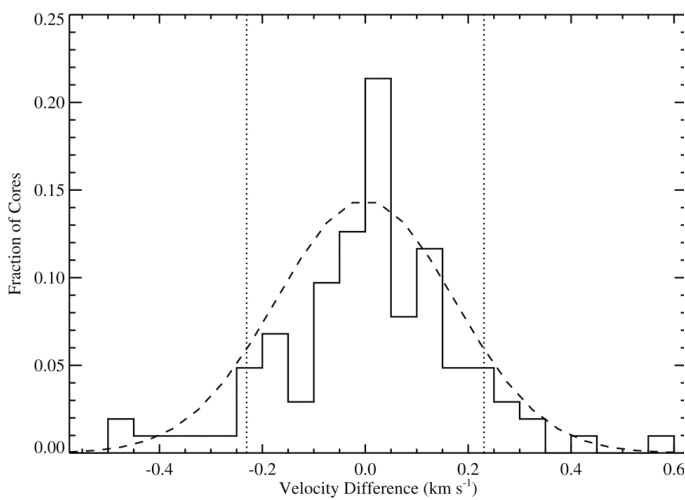

*Figure 5 — The relative motion between the dense core and its surrounding envelope. The vertical dotted lines indicate the local sound speed, ~0.23 km s$^{-1}$. The dashed line shows a Gaussian fit to the histogram; the standard deviation of the observations is 0.17km s$^{-1}$. Adapted from Kirk, Johnstone, & Tafalla (2007).*

Perseus dense cores versus simulated observations of two turbulence-dominated simulations tuned to have large-scale properties similar to Perseus (Offner *et al.* 2008). In one of the simulations, the turbulence was continuously driven on large scales, while in the other, it was allowed to decay as the simulation evolved. Our observations clearly rule out the driven simulation – there are far too many cores predicted that have large non-thermal motions. (We saw no cores with $\sigma_{nt} / c_S$ above 1.8). The decaying turbulence scenario (solid light-blue line) is more consistent, although even there, too many dense cores are predicted at high $\sigma_{nt} / c_S$. The distribution of non-thermal motions within dense cores therefore provides an important avenue by which to test simulations.

### 2.2.2 Small Core-Envelope Motion

A second measure that provides a strong test for turbulent simulations is the motion between the dense core and its surrounding envelope. Walsh, Myers, & Burton (2004; hereafter WMB04) argued that turbulent simulations in which dense cores form via competitive accretion should show large relative motions between the dense cores and their envelopes. This was contrary to their observations of small motions between dense cores and their envelopes, although their survey focussed on dense cores in relatively isolated environments. Following WMB04, Ayliffe *et al.* (2007) made mock observations of a competitive accretion simulation, and found that they could reproduce the WMB04 results under certain circumstances, although in this case, the cores were not simultaneously thermally dominated. Unlike isolated star-forming environments, competitive accretion is expected to dominate in more clustered conditions, and our survey (Paper II) was the first to make similar measurements in this regime; see Figure 5. Similar to WMB04, we found that the core-envelope motion was small, usually less than the sound speed, with a standard deviation of 0.17 km s$^{-1}$, or ~75 percent of the sound speed. Debate continues as to whether competitive accretion simulations can match these observations. For example, in their mock observations of a competitive accretion simulation, Rundle *et al.* (2010) found small core-envelope motions with a standard deviation of 0.16 km s$^{-1}$, but also had core (N$_2$H$^+$) linewidths that were consistently non-thermal and *larger* than the corresponding envelope (C$^{18}$O) linewidths (see their Figure 17). Clearly more work is still needed to investigate these simulations.

### 2.2.3 Small Core-Core Motion

As a separate component of the COMPLETE survey, a full $^{13}$CO(1-0) map of the Perseus cloud was made using the FCRAO (Ridge *et al.* 2006, Pineda, Caselli, & Goodman 2008). $^{13}$CO is an excellent tracer of the bulk material within a molecular cloud, and so provides a complement to the KJT07 pointed dense-gas spectral survey. In Kirk *et al.* (2010, hereafter Paper III), we used the large-scale structures "extinction regions" identified in KJD06 to define sub-regions of Perseus to study.

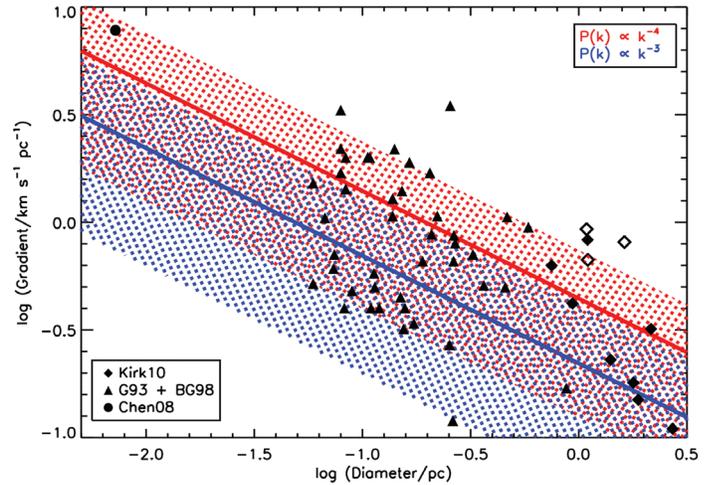

*Figure 6 — The velocity gradient measured in $^{13}$CO in each of the extinction regions versus the region size (diamonds). Regions where the measurement was more uncertain, due to incomplete coverage in $^{13}$CO, are denoted by the open diamonds. Observations of NH3 cores from Goodman* et al. *(1993) and Barranco & Goodman (1998) are shown by the triangles, and a high-resolution measurement of BHR-71 by Chen* et al. *(2008) is shown by the circle. The red and blue shading and lines show the zone and most likely values predicted for observations of regions dominated by large-scale modes of turbulence by Burkert & Bodenheimer (2000) for power spectra of P(k) ∝ k$^{-3}$ and k$^{-4}$ respectively. Adapted from Kirk* et al. *(2010).*

Most of the extinction regions corresponded to well-known star-forming regions such as IC348 and NGC1333. Within each extinction region, we measured the core-to-core velocity dispersion, as well as the total $^{13}$CO velocity dispersion over the same region (calculated by summing all the $^{13}$CO spectra). We found that the core-to-core velocity dispersions were roughly half as large as the $^{13}$CO gas velocity dispersions. The number of cores in each extinction region was usually small, and the cores were often clustered within a small part of the region, however, neither of these effects appeared to be responsible for the factor of two difference in the velocity dispersions measured. Interpretation depends on the model adopted for the three-dimensional structure of the cloud (*i.e.* how much substructure there is along the line of sight), however, the measurement is straightforward to apply to simulations where the full cloud structure is known, and offers another way to test simulations.

### 2.2.4 Bulk Cloud Motions

Finally, in Paper III we analyzed the large-scale kinematic properties of the $^{13}$CO gas. We measured the velocity gradient across each extinction region following the procedure outlined in Goodman *et al.* (1993). Gas following a turbulent power spectrum dominated by the largest-scale modes, as is usually assumed in turbulent simulations (*e.g.* Offner *et al.* 2008, Rundle *et al.* 2010), should follow a well-defined relationship between the velocity gradient and the size of the region for large-scale regions (Burkert & Bodenheimer 2000). Figure 6 shows the velocity gradient observed versus the diameter of each region (diamonds). The relationship and spread in observed values predicted for a turbulent power spectrum of P ∝ $k^{-4}$ and $k^{-3}$ by Burkert & Bodenheimer (2000) are shown by the red and blue lines and shaded regions. The classical Kolmogorov incompressible turbulent power spectrum has a slope of -11/3, which lies between these values. For comparison, observations from Goodman *et al.* (1993) and Barranco & Goodman (1998) of $NH_3$ cores are shown by triangles, and a recent high-resolution observation of a dense core in the isolated Bok globule, BHR 71 from Chen *et al.* (2008) is shown by the circle. Clearly, the observations all follow the same trend and agree with the prediction for a turbulent power spectrum of roughly $k^{-4}$. Turbulence dominated by large modes therefore provides a good description of this cloud property.

## 3. Simulations

As can be seen from above, fully interpreting the observations usually requires comparison with predictions from simulations, which ideally are "observed" in a similar fashion as the actual observations. Few simulations yet incorporate magnetic fields (although one notable exception is the work of Nakamura & Li 2008), and most are run with a single set of initial conditions, due to computational limitations. The interaction between magnetic fields and turbulent motions and their combined effect on the evolution of the cloud is non-trivial, making parameter studies an important, but so far under-utilized approach. For that reason, we (Kirk, Johnstone, & Basu 2009) analyzed a suite of 21 simulations with varying initial magnetic-field strengths and levels of turbulence. These simulations were based on the setup described in Basu, Ciolek, & Wurster (2009), Ciolek & Basu (2006), and references therein. The simulations we analyzed had input Mach numbers ranging from 0 (thermal motions only) to 4 (dominated by turbulence) and an initial magnetic field ranging from very strong ($\mu_0$ = 0.5) to very weak ($\mu_0$ = 2), where $\mu_0$ is the initial mass to magnetic-flux ratio. In order for the simulations to be able to run in a reasonable amount of time, they had a thin-sheet geometry, rather than being fully 3-D, however, they did include non-ideal MHD effects. Briefly, "observations" of these simulations showed that the distribution of core linewidths (2.2.1) was reproduced reasonably well in all of the simulations, however, the core-envelope motion (2.2.2) was more difficult – see Figure 7. None of the simulations we analyzed showed both small core-envelope motions and significant non-thermal velocity dispersion on the large scale (as observed in $^{13}CO$ for the extinction regions). This discrepancy with the observations may be due to the boundary conditions or geometry of the simulations, or indicate that additional physical processes are important. The analysis more importantly demonstrated that the core-envelope motion was an excellent discriminant for star-formation models, as we found it depends strongly on the initial conditions adopted in simulations.

## 4. Outlook for the Future

The observational benchmarks highlighted here can be used in conjunction with synthetic observations of numerical simulations in order to determine which processes dominate star formation both on the cloud-wide and dense core scale. Numerical simulations are ever improving to include more physical processes, a finer level of detail, or span a larger parameter-space, and observations also face an exciting future. The strength of the observational measures presented here is the cloud-wide coverage in a variety of tracers, which has only recently become possible. So far, this type of analysis has not been applied to many clouds. That situation is changing, however, with Gould Belt Legacy Surveys (GLBS) in progress or completed at several telescopes, including *Spitzer*, *Herschel*, and the JCMT. The Gould Belt covers all nearby major star-forming regions, and hence, in the near future, measures similar to those discussed here can be made on a variety of molecular clouds. With this information, we will be able to understand the influence of environment on star-formation processes, and learn how typical Perseus is. Canadians have a strong involvement in the JCMT GBLS (Ward-Thompson *et al.* 2007), particularly the continuum-mapping component with the newly developed SCUBA-2 array. SCUBA-2 features a larger, more-sensitive detector than its predecessor, SCUBA, with an expected increase in mapping speed of well above an order of magnitude, making it ideal for large mapping projects. We are excited to see SCUBA-2 now in the commissioning phase at the JCMT.

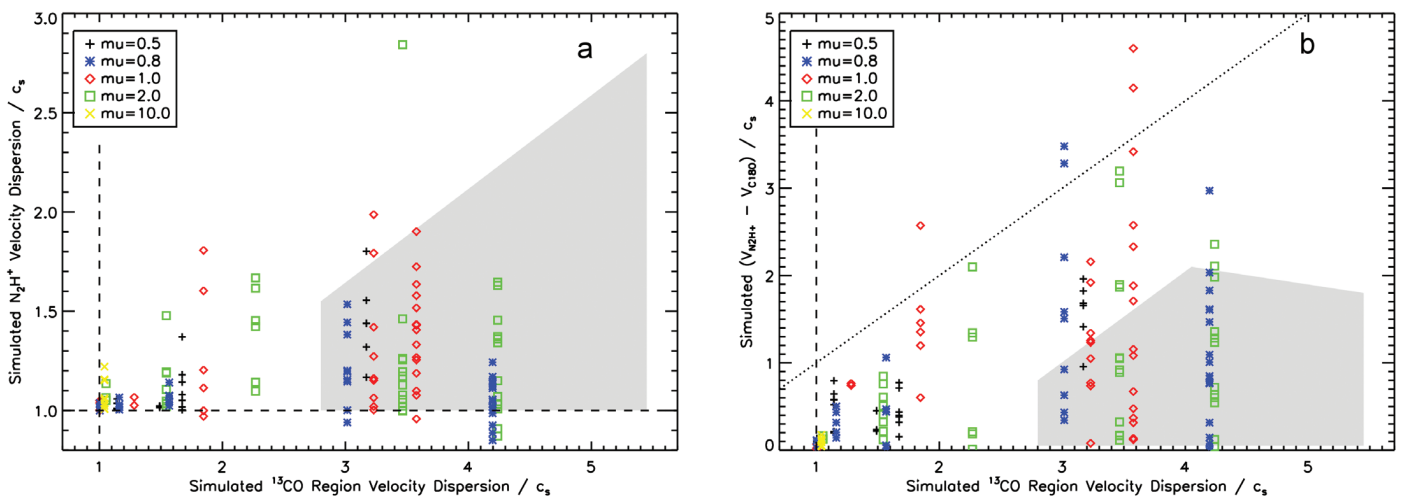

*Figure 7 — A comparison between simulations and observations, with different initial magnetic-field strengths in the simulations denoted by different colours. The range spanned by the observations is shown by the grey shading. Left panel (a): velocity dispersion of $N_2H^+$ (the dense core gas) versus that of $^{13}CO$ (the cloud gas) across the entire extinction region or simulated box, in units of the sound speed. The dashed lines denote the sound speed, which is the minimum value expected. All simulations show a similar range in $N_2H^+$ velocity dispersion, independent of the region velocity dispersion, and the observations show a similar range. Right panel (b): The velocity difference between $N_2H^+$ and $C^{18}O$ (core and envelope gas in our observations) versus the $^{13}CO$ velocity dispersion of the extinction region or simulated box. The dotted line denotes a 1-1 correspondence. The observations show a much smaller range than the simulations at similar values of the $^{13}CO$ velocity dispersion. Adapted from Kirk, Johnstone, & Basu (2009).*

Complementary to the developments in large-scale surveys are major advances for small-scale, high-resolution studies. The Atacama Large Millimetre Array (ALMA) is an array of more than 60 submillimetre dishes located in the Atacama desert at an altitude of 5000 m in Northern Chile. This facility represents a substantial increase in capabilities over current interferometric facilities, and will be operated by an international consortium including partners from North America, Europe, and East Asia. Canada is a major partner in the North American group and, through the instrument group at the Herzberg Institute for Astrophysics (in Victoria), is providing one of the key receivers for the dishes. ALMA will have the capability of studying structures at sub-arcsecond resolution for continuum or multiple spectral lines simultaneously. Many avenues of exploration will be opened by this new facility, including the fragmentation of dense cores (examining the formation of multiple-star systems) and the radial distribution of molecules within dense cores (tracing the chemical processes at work). Early science observations are expected to commence in less than a year.

## 5. Conclusions

The wealth of data covering the Perseus molecular cloud has allowed an unprecedented study of the cloud properties and the development of a series of statistically significant observational constraints on the processes shaping the evolution of the molecular cloud and dense cores within it. Several of these were highlighted here, summarizing the work in Kirk *et al.* (2006, 2007, 2009, 2010). A fuller, though likely still incomplete, list of studies based on recent Perseus survey data includes Curtis *et al.* (2010a, 2010b), Enoch *et al.* (2006), Foster *et al.* (2006, 2008, 2009), Goodman, Pineda, & Schnee (2009), Hatchell *et al.* (2005, 2007a, 2007b, 2008, 2009), Johnstone *et al.* (2010), Jorgensen *et al.* (2006, 2007, 2008), Kauffmann *et al.* (2010), Pineda *et al.* (2008, 2009, 2010), and Rosolowsky *et al.* (2008).

The results highlighted here include:

- Dense cores are only found in regions where the large-scale column density corresponds to a visual extinction of ~5 magnitude or higher. This is in agreement with predictions from a magnetically dominated formation scenario.

- Dense cores are thermally dominated, with only a small fraction (8 percent of starless cores) having a ratio of non-thermal to thermal motions greater than 1. This appears to rule out some models of turbulence-dominated formation.

- The motion between dense cores and their surrounding envelopes is small, typically less than the sound speed. There is debate over whether the competitive accretion scenario within a turbulence-dominated model can match this observation.

- The motion between dense cores within larger structures (extinction regions) is also small, roughly half of the total $^{13}$CO gas velocity dispersion within the region. This measurement can be easily applied to future star-formation simulations.

- In each extinction region, the relationship between the $^{13}$CO velocity gradient and dispersion is similar to that predicted for a turbulent power spectrum dominated by large-scale modes, with $P(k) \propto k^{-4}$.

The multi-wavelength coverage of Perseus has led to a richer understanding of this star-forming environment. Further advances will be made following upcoming surveys of other clouds. The next few years in star-formation research promise to provide substantial advances on the questions now posed, and will undoubtedly raise exciting new puzzles as well. Stay tuned!

## Acknowledgements

HK thanks her many collaborators in the research highlighted here, in particular, her thesis advisor Doug Johnstone (HIA/UVic), as well as Shantanu Basu (UWO), James Di Francesco (HIA/UVic), Alyssa Goodman (CfA), Jaime Pineda (Manchester), and Mario Tafalla (OAN). HK also thanks Doug Johnstone for providing useful suggestions about this manuscript. HK is supported by a Natural Sciences and Engineering Research Council of Canada Postdoctoral Fellowship, with additional support from the Smithsonian Astrophysical Observatory. She gratefully acknowledges support from NSERC, NRC, and University of Victoria fellowships during her thesis research. ✶

# The Royal Astronomical Society of Canada

*Vision* — To inspire curiosity in all people about the Universe, to share scientific knowledge, and to foster collaboration in astronomical pursuits.

*Mission* — The Royal Astronomical Society of Canada (RASC) encourages improved understanding of astronomy for all people, through education, outreach, research, publication, enjoyment, partnership, and community.

*Values* — The RASC has a proud heritage of excellence and integrity in its programmes and partnerships. As a vital part of Canada's science community, we support discovery through the scientific method. We inspire and encourage people of all ages to learn about and enjoy astronomy.